\documentstyle[twoside,fleqn,espcrc2,epsfig]{article}

\def\lsim{\raise0.3ex\hbox{$<$\kern-0.75em\raise-1.1ex\hbox{$\sim$}}}
\def\gsim{\raise0.3ex\hbox{$>$\kern-0.75em\raise-1.1ex\hbox{$\sim$}}}

\title{The Strong-Coupling Expansion in Simplicial Quantum Gravity}

\author{
  S.~Bilke\address{Inst.~Theor.~Fysica, Univ.~Amsterdam, 
                   1018 XE Amsterdam, The Netherlands \\
		   $^{\rm b}$Institute of Physics, Jagellonian
                   University, 30059 Krakow, Poland \\
		   $^{\rm c}$LPTHE, B\^{a}timent 211, Universit\'e 
                   Paris-Sud, 91405 Orsay, France \\
		   $^{\rm d}$Fakult\"{a}t f\"{u}r Physik, Universit\"{a}t 
                   Bielefeld, 33501 Bielefeld, Germany }
  Z.~Burda$^{\rm b}$, 
  A.~Krzywicki$^{\rm c}$, 
  {\bf B.~Petersson}$^{\rm d}$,
  K.~Petrov$^{\rm d}$,
  J.~Tabaczek$^{\rm d}$ 
  and G.~Thorleifsson$^{\rm d}$
}		             

\begin{document}

\begin{abstract}
We construct the strong-coupling series in 4d simplicial
quantum gravity up to volume 38. It is used to calculate
estimates for the string susceptibility exponent $\gamma$
for various modifications of the theory. It provides a 
very efficient way to get a first view of the phase structure 
of the models.
\end{abstract}

\maketitle

\section{INTRODUCTION}

Euclidean quantum gravity is formally defined by 
the partition function
\begin{equation}
 Z \;=\; \int\,\frac{D [g_{\mu\nu}]}{\rm Vol(diff)} 
 \;{\rm e}^{\textstyle \,- S[g_{\mu\nu}]}
 \label{part.cont}
\end{equation}
where the Einstein-Hilbert action
\begin{equation}
 S\left[g_{\mu\nu}\right] \;=\; 
 \int\, {\rm d}^4 \xi \sqrt{g (\xi)} \left\{
 \lambda - \frac{1}{16\pi G} R (\xi) \right\}.
 \label{act.cont}
\end{equation}
Discretizing the theory on an ensemble of four-dimensional 
manifolds $T$, consisting of $N_4$ 4-simplexes glued together 
along the faces, and assuming a metric where all link lengths 
are equal to one, leads to simplicial quantum gravity:
\begin{equation}
 Z_{GC}(\kappa_2,\kappa_4) \;=\; \sum_{\{T\} } \,
 \frac{1}{C(T)} \, {\rm e}^{\; \kappa_2 N_2 (T) - \kappa_4 N_4 (T)}.
 \label{part.grand}
\end{equation}
Here $\kappa_4$ corresponds to the cosmological constant
$\lambda$, $\kappa_2$ to the inverse Newton's constant $G$, 
and $N_2$ is the number of 2-simplexes (triangles). $C(T)$ is
the symmetry factor of triangulation $T$, the number of 
equivalent labelings of the vertexes.
Eq.~(\ref{part.grand}) is related to
the canonical (fixed volume) partition function
$Z_{N_4}$ through
\begin{equation}
 Z_{GC}(\kappa_2,\kappa_4) \;=\; \sum_{N_4} \, {\rm e}^{-\kappa_4 N_4}
 \; Z_{N_4}(\kappa_2).
 \label{part.canonical}
\end{equation}

Although it is not proven analytically, one assumes that
$Z_{N_4}$ is exponentially bounded in $N_4$.
Moreover, in our analysis we assume
that 
\begin{equation}
 Z_{N_4}(\kappa_2) \;\sim\; N_4^{\gamma(\kappa_2)-3} \,
 {\rm e}^{\mu_c (\kappa_2) N_4} \, 
 (1 + {\scriptstyle \frac{c_1}{N_4}} + ...)
 \label{asymptotic}
\end{equation}
when $N_4 \rightarrow \infty$, where $\gamma$ determines 
the critical behavior of the grand-canonical partition function:
$Z_{GC} \sim (\mu_c-\kappa_4)^{2-\gamma}$ as 
$\kappa_4 \rightarrow \mu_c$.

Simulating this model one observes
two phases; a branched polymer phase for
$\kappa_2 \,\gsim\, 1.3$, and a crumpled phase for 
$\kappa_2 \,\lsim\, 1.3$. 
In the branched polymer phase $\gamma=\frac{1}{2}$,
whereas in the crumpled phase the sub-leading
correction is exponential, rather than Eq.~(\ref{asymptotic});
this corresponding formally to $\gamma= - \infty$. 
Separating the two phases is a first order phase transition,
hence no interesting critical behavior is observed. 
For that one may need to modify the
simple action in Eq.~(\ref{part.grand}) above. However, as 
investigating  every modification requires extensive numerical 
simulations, it is important to have some analytical guidance. 
One such is the strong-coupling expansion.

We have used this expansion to investigate two modifications of 
the model, described in Ref.~\cite{bbkptt1}, 
namely the interaction of gravity with gauge matter fields, 
and a modified measure.

\section{STRONG-COUPLING EXPANSION}

The strong-coupling expansion consist of explicit 
construction of the smallest configurations and their 
symmetry factors, and a calculation of the additional 
weights corresponding to a given model.  
In two dimensions matrix models allow a recursive construction
of the series; in four dimensions, though,
such methods are not available, but an alternative procedure
based on numerical simulations has been developed
\cite{bbkptt1}:

\vspace{4pt}
\noindent
({\sl a}) Using Monte Carlo simulations we identify all
{\it distinct} triangulations, up to a given volume.
To identify the triangulations, we introduce a
hash function $f(T)$; this function has
to be complicated enough to distinguish between
combinatorially different triangulations, while simple
enough for a repeated calculation in the MC
simulations.  

Only triangulations that can not be reduced by
applying one of the volume decreasing geometric
moves have to be considered in the MC.
All others are systematically constructed from smaller 
ones.  This reduces the MC efforts
substantially as, in practice, 99.9\% of 
the triangulations turn out to be reducible.

\vspace{4pt}
\noindent
({\sl b}) For each triangulation we calculate the
corresponding symmetry factor $C(T)$ by comparing
all possible permutations of the vertex labels
(only permutations of vertexes
 of same order have to be considered).
Comparing the $C(T)$'s to the relative
frequency with which different triangulations appear
in the MC simulations provides a
check on the correct identification
of the triangulations.

\vspace{4pt}
Using this procedure we have identified all four-dimensional
triangulations up to volume $N_4 = 38$, all in all
1.477.713 distinct triangulations.  This gives the first
16 terms in the strong-coupling series, however,
due to different finite-size corrections these terms
split into two distinct series which have to be analyzed
separately.

The strong-coupling series is analyzed using an
appropriate series extrapolation method, e.g.\ the
{\it ratio method}.  Assuming the asymptotic
behavior Eq.~(\ref{asymptotic}) this yields the
critical coupling $\mu_c$ and the 
exponent $\gamma$.

\begin{figure}[t]
 \begin{center}
  \psfig{figure=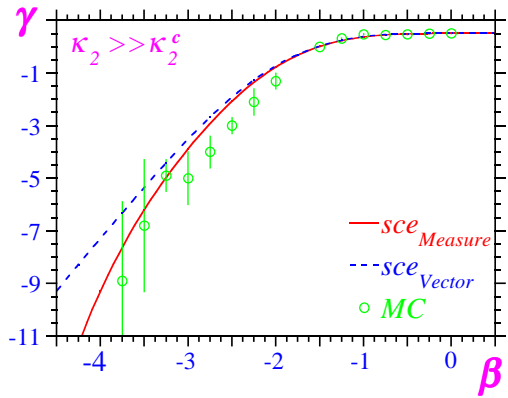,width=2.6in}
 \end{center}
 \label{fig1}
 \caption{Strong-coupling estimates of $\gamma$ for both
  a modified measure (solid line) and $f$ copies
  of vector fields (dashed line). The circles are
  values of $\gamma$ from MC with modified measure.}
\end{figure}

\section{RESULTS}

Using the strong-coupling expansion it is easy to explore a large
set of modified models of 4d simplicial gravity. 
Here we will give results just for two such models:

\vspace{2pt}
\noindent
({\sl a}) Modified measure:
\vspace{-2pt}
\begin{equation}
 M(T) \;\sim\; \prod_{j=1}^{N_2} o(t_j)^\beta
 \label{part.measure}
\end{equation}
\vspace{-5pt}
where $o(t_j)$ is the order of the $j$'th triangle, i.e.\ 
to how many 4-simplexes it is attached.

\noindent
({\sl b}) $f$ copies of vector fields:
\vspace{-5pt}
\begin{equation}
 Z_V(T) \, = \, \int' \prod_{l \in T} {\rm d}A(l) \; 
 e^{- S[A(l)]}
 \label{part.vector}
\end{equation}
\vspace{-10pt}
\[
 S(A(l)) \;=\; \sum_{t_{abc}} o(t_{abc}) \left[ A(l_{ab}) + A(l_{bc})
 +A(l_{ca})\right]^2
 \nonumber
\]
\vspace{-3pt}
Here $A(l)$ are non-compact $U(1)$ gauge fields residing on 
the links and the sum is over all triangles.
The prime indicates that the zero modes of the gauge field
are not integrated over. 

The weights corresponding to both models are calculated
explicitly for every configuration $T$. 
The series so obtained are analyzed with the ratio method, 
assuming the asymptotic behavior Eq.~(\ref{asymptotic}). 
In Fig.~1 we shown an example of $\gamma$, for 
$\kappa_2 \geq \kappa_2^c$, for both the models
as we vary $\beta$ and $f$ respectively. 
For consistency we have verified that successive
approximations in the analysis of the series converge, 
what is plotted is the result of the highest approximation.
These results are in excellent agreement with values 
of $\gamma$ measured in MC simulations 
of Eq.~(\ref{part.measure}) for $N_4=4000$.
Furthermore, the two models yield the same
variation of $\gamma$ provided one takes
\begin{equation}
 \beta \;=\; - \frac{f}{2} - \frac{1}{4}.
\end{equation}
This universality appears in a variety of  
``reasonable'' modifications of the model Eq.~(\ref{part.grand}),
all of which give the same phase structure 
(modulo a trivial re-scaling of parameters) \cite{bbkptt1}.

\begin{figure}[t]
 \begin{center}
  \psfig{figure=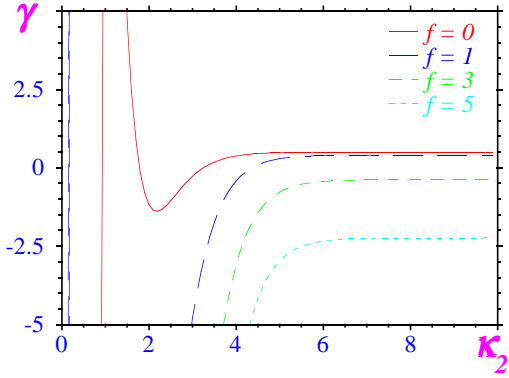,width=2.6in}
 \end{center}
 \label{fig2}
 \caption{Variations in $\gamma$ {\it versus} $\kappa_2$
  for $f$ copies of vector fields coupled to 4d--gravity.}
\end{figure}

\begin{figure}[t]
 \psfig{figure=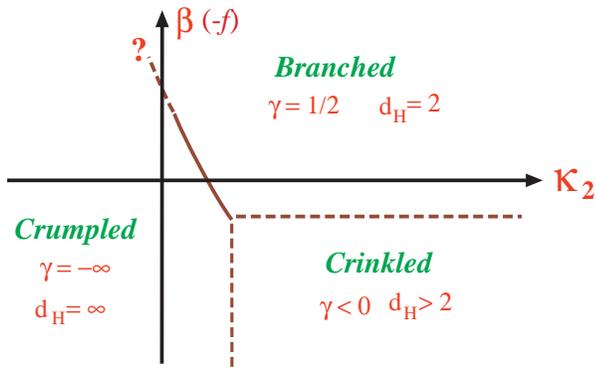,width=2.8in}
 \label{fig3}
 \caption{A schematic phase diagram of simplical gravity
  coupled to either a measure term, or to vector fields.}
\end{figure}

Fig.~2 shows $\gamma$ as we vary $\kappa_2$,
for different number of vector fields, given by the
strong-coupling series.  The values of  
$\gamma$ become unstable for small $\kappa_2$; 
this is consistent with the fact that
in the corresponding MC simulations we observe a
phase transition to a crumpled phase. 
This transition is evident in the series extrapolation
as successive approximations fail to converge both
in the crumpled phase and in the critical region.

From these and further investigations we conclude that phase
diagrams of the two models are quite similar, 
as depicted schematically in Fig.~3.  In addition to
the crumpled and branched polymer phases, a
new {\it crinkled} phase appears, for sufficiently
strong coupling to either matter or the measure
term,  characterized by a finite $\gamma < 0$. 

As $\kappa_2 \rightarrow \infty$, the
finite-size effects become, rather
surprisingly, less important in the pure 
gravity model.  This is related to the dominance of a
particular class of manifolds, {\it stacked spheres}, in 
this limit \cite{gab}.   However, as a coupling to
a measure term is added the finite-size effects increase
again, as shown in Fig.~4 were we plot the  
the first correction term to Eq.~(\ref{asymptotic})
for $\kappa_2 = \infty$ and varying $\beta$.
The increased finite-size correction coincided with
the appearance of the crinkled phase.

\begin{figure}[t]
 \begin{center}
 \psfig{figure=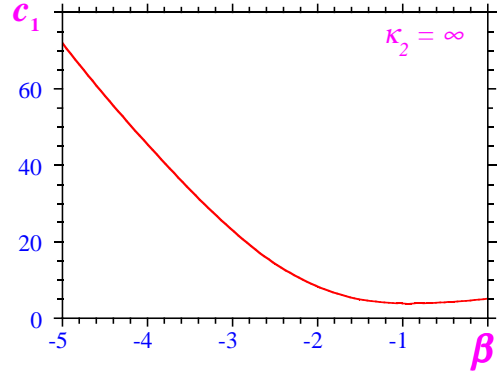,width=2.6in}
 \end{center}
 \label{fig4}
 \caption{The first corrections term in Eq.~(5).}
\end{figure}

The examples given above show that the strong-coupling series
gives a good qualitative view of the phase structure of
modified models of 4d simplicial gravity. It provides us with 
a powerful tool and should be used before
starting extensive computer simulations of a particular model.


\begin{thebibliography}{99}

\bibitem{first}
 P.~Bialas, Z.~Burda, A.~Krzywicki and B.~Petersson,
  {\it Nucl.~Phys.~B} {\bf 472} (1996) 293.

\bibitem{bbkptt1} 
 S.~Bilke, Z.~Burda, A.~Krzywicki, B.~Petersson,
 J.~Tabaczek, G.~Thorleifsson, 
 {\it Phys.\ Lett.~B} {\bf 418} (1998) 266;
 {\it ibid}. {\bf 432} (1998) 279. 

\bibitem{gab} 
 D.~Gabrielli, {\it Phys.~Lett.~B} {\bf 421} (1998) 79.
\end{thebibliography}
\end{document}